%
%
\magnification=\magstep1
\advance\voffset by 3truemm   
\advance\hoffset by -1truemm   
\vsize=23truecm  \hsize=16.5truecm
\overfullrule=0pt
\hfuzz 15truept
\parskip=5pt
\baselineskip=12pt
\font\tit=cmbx10 scaled \magstep2
\font\subt=cmbx10 scaled \magstep1
\font\ssubt=cmbx10
\font\auth=cmr10 scaled \magstep1
\font\titabs=cmti9
\font\abs=cmr9
\def\title#1{\null\vskip 26truemm \noindent {\tit #1} \vskip 12truemm 
              \noindent {\auth By} } 
\def\tbreak{\hfil\vskip 0truemm\noindent}  
\def\authors#1{\noindent {\auth #1}   \vskip 5truemm}
\def\address#1{\noindent #1\vskip 7truemm}
\def\abstract#1{\vskip 19truemm\noindent 
   {\abs {\titabs Abstract} #1  }} 
\def\section#1{\vskip 12truemm \noindent{\tit #1}\vskip 5mm\noindent}
\def\subsection#1{\vskip 8truemm \noindent{\subt #1}\vskip 3truemm\noindent}
\def\subsubsection#1{\vskip 5truemm \noindent{\ssubt #1}\vskip 2truemm\noindent}


\input epsf


\def\ts{{\thinspace}}
\def\ref#1{\lbrack #1\rbrack}

\def\part#1#2{{\partial #1\over\partial #2}}

\def\Real{{\rm I\mathchoice{\kern-0.70mm}{\kern-0.70mm}{\kern-0.65mm}%
  {\kern-0.50mm}R}}
\def\C{\rm C\kern-.42em\vrule width.03em height.58em depth-.02em
       \kern.4em}

\def\bx#1{\leavevmode\thinspace\hbox{\vrule\vtop{\vbox{\hrule\kern1pt
        \hbox{\vphantom{\tt/}\thinspace{\bf#1}\thinspace}}
      \kern1pt\hrule}\vrule}\thinspace}

{\catcode`\@=11
\gdef\SchlangeUnter#1#2{\lower2pt\vbox{\baselineskip 0pt \lineskip0pt
  \ialign{$\m@th#1\hfil##\hfil$\crcr#2\crcr\sim\crcr}}}
}

\def\ueber#1#2{{\setbox0=\hbox{$#1$}%
  \setbox1=\hbox to\wd0{\hss$\scriptscriptstyle #2$\hss}%
  \offinterlineskip
  \vbox{\box1\kern0.4mm\box0}}{}}

\def\bx#1{\leavevmode\thinspace\hbox{\vrule\vtop{\vbox{\hrule\kern1pt
        \hbox{\vphantom{\tt/}\thinspace{\bf#1}\thinspace}}
      \kern1pt\hrule}\vrule}\thinspace}

\input epsf

%
%

\font\kleinhalbcurs=cmmib10 scaled 833
\font\eightrm=cmr8
\font\sixrm=cmr6
\font\eighti=cmmi8
\font\sixi=cmmi6
\skewchar\eighti='177 \skewchar\sixi='177
\font\eightsy=cmsy8
\font\sixsy=cmsy6
\skewchar\eightsy='60 \skewchar\sixsy='60
\font\eightbf=cmbx8
\font\sixbf=cmbx6
\font\eighttt=cmtt8
\hyphenchar\eighttt=-1
\font\eightsl=cmsl8
\font\eightit=cmti8

\font\bxf=cmbx10
  \mathchardef\Gamma="0100
  \mathchardef\Delta="0101
  \mathchardef\Theta="0102
  \mathchardef\Lambda="0103
  \mathchardef\Xi="0104
  \mathchardef\Pi="0105
  \mathchardef\Sigma="0106
  \mathchardef\Upsilon="0107
  \mathchardef\Phi="0108
  \mathchardef\Psi="0109
\def\rahmen#1{\vskip#1truecm}
\def\begfig#1cm#2\endfig{\par
\setbox1=\vbox{\rahmen{#1}#2}%
\dimen0=\ht1\advance\dimen0by\dp1\advance\dimen0by5\baselineskip
\advance\dimen0by0.4true cm
\ifdim\dimen0>\vsize\pageinsert\box1\vfill\endinsert
\else
\dimen0=\pagetotal\ifdim\dimen0<\pagegoal
\advance\dimen0by\ht1\advance\dimen0by\dp1\advance\dimen0by1.4true cm
\ifdim\dimen0>\vsize
\topinsert\box1\endinsert
\else\vskip1true cm\box1\vskip4true mm\fi
\else\vskip1true cm\box1\vskip4true mm\fi\fi}
\def\figure#1#2{\smallskip\setbox0=\vbox{\noindent\petit{\bf Fig.\ts#1.\
}\ignorespaces #2\smallskip
\count255=0\global\advance\count255by\prevgraf}%
\ifnum\count255>1\box0\else
\centerline{\petit{\bf Fig.\ts#1.\ }\ignorespaces#2}\smallskip\fi}


\def\begtab#1cm#2\endtab{\par
\ifvoid\topins\midinsert\vbox{#2\rahmen{#1}}\endinsert
\else\topinsert\vbox{#2\kern#1true cm}\endinsert\fi}
\def\rahmen#1{\vskip#1truecm}
\def\begpet{\vskip6pt\bgroup\petit}
\def\endpet{\vskip6pt\egroup}
\def\begref{\par\bgroup\petit
\let\it=\rm\let\bf=\rm\let\sl=\rm\let\INS=N}
\def\petit{\def\rm{\fam0\eightrm}%
\textfont0=\eightrm \scriptfont0=\sixrm \scriptscriptfont0=\fiverm
 \textfont1=\eighti \scriptfont1=\sixi \scriptscriptfont1=\fivei
 \textfont2=\eightsy \scriptfont2=\sixsy \scriptscriptfont2=\fivesy
 \def\it{\fam\itfam\eightit}%
 \textfont\itfam=\eightit
 \def\sl{\fam\slfam\eightsl}%
 \textfont\slfam=\eightsl
 \def\bf{\fam\bffam\eightbf}%
 \textfont\bffam=\eightbf \scriptfont\bffam=\sixbf
 \scriptscriptfont\bffam=\fivebf
 \def\tt{\fam\ttfam\eighttt}%
 \textfont\ttfam=\eighttt
 \normalbaselineskip=9pt
 \setbox\strutbox=\hbox{\vrule height7pt depth2pt width0pt}%
 \normalbaselines\rm
\def\vec##1{\setbox0=\hbox{$##1$}\hbox{\hbox
to0pt{\copy0\hss}\kern0.45pt\box0}}}%
\let\ts=\thinspace
%
\font \tafontt=     cmbx10 scaled\magstep2
\font \tafonts=     cmbx7  scaled\magstep2
\font \tafontss=     cmbx5  scaled\magstep2
\font \tamt= cmmib10 scaled\magstep2
\font \tams= cmmib10 scaled\magstep1
\font \tamss= cmmib10
\font \tast= cmsy10 scaled\magstep2
\font \tass= cmsy7  scaled\magstep2
\font \tasss= cmsy5  scaled\magstep2
\font \tasyt= cmex10 scaled\magstep2
\font \tasys= cmex10 scaled\magstep1
\font \tbfontt=     cmbx10 scaled\magstep1
\font \tbfonts=     cmbx7  scaled\magstep1
\font \tbfontss=     cmbx5  scaled\magstep1
\font \tbst= cmsy10 scaled\magstep1
\font \tbss= cmsy7  scaled\magstep1
\font \tbsss= cmsy5  scaled\magstep1

\newbox\chsta\newbox\chstb\newbox\chstc
\def\centerpar#1{{\advance\hsize by-2\parindent
\rightskip=0pt plus 4em
\leftskip=0pt plus 4em
\parindent=0pt\setbox\chsta=\vbox{#1}%
\global\setbox\chstb=\vbox{\unvbox\chsta
\setbox\chstc=\lastbox
\line{\hfill\unhbox\chstc\unskip\unskip\unpenalty\hfill}}}%
\leftline{\kern\parindent\box\chstb}}
 \def \chap#1{
    \vskip24pt plus 6pt minus 4pt
    \bgroup
 \textfont0=\tafontt \scriptfont0=\tafonts \scriptscriptfont0=\tafontss
 \textfont1=\tamt \scriptfont1=\tams \scriptscriptfont1=\tamss
 \textfont2=\tast \scriptfont2=\tass \scriptscriptfont2=\tasss
 \textfont3=\tasyt \scriptfont3=\tasys \scriptscriptfont3=\tenex
     \baselineskip=18pt
     \lineskip=18pt
     \raggedright
     \pretolerance=10000
     \noindent
     \tafontt
     \ignorespaces#1\vskip7true mm plus6pt minus 4pt
     \egroup\noindent\ignorespaces}%
 \def \sec#1{
     \vskip25true pt plus4pt minus4pt
     \bgroup
 \textfont0=\tbfontt \scriptfont0=\tbfonts \scriptscriptfont0=\tbfontss
 \textfont1=\tams \scriptfont1=\tamss \scriptscriptfont1=\kleinhalbcurs
 \textfont2=\tbst \scriptfont2=\tbss \scriptscriptfont2=\tbsss
 \textfont3=\tasys \scriptfont3=\tenex \scriptscriptfont3=\tenex
     \baselineskip=16pt
     \lineskip=16pt
     \raggedright
     \pretolerance=10000
     \noindent
     \tbfontt
     \ignorespaces #1
     \vskip12true pt plus4pt minus4pt\egroup\noindent\ignorespaces}%
 \def \subs#1{
     \vskip15true pt plus 4pt minus4pt
     \bgroup
     \bxf
     \noindent
     \raggedright
     \pretolerance=10000
     \ignorespaces #1
     \vskip6true pt plus4pt minus4pt\egroup
     \noindent\ignorespaces}%
 \def \subsubs#1{
     \vskip15true pt plus 4pt minus 4pt
     \bgroup
     \bf
     \noindent
     \ignorespaces #1\unskip.\ \egroup
     \ignorespaces}
\def\footnoterule{\kern-3pt\hrule width 2true cm\kern2.6pt}
\newcount\footcount \footcount=0
\def\advftncnt{\advance\footcount by1\global\footcount=\footcount}
\def\fonote#1{\advftncnt$^{\the\footcount}$\begingroup\petit
       \def\textindent##1{\hang\noindent\hbox
       to\parindent{##1\hss}\ignorespaces}%
\vfootnote{$^{\the\footcount}$}{#1}\endgroup}

\newcount\sterne
\outer\def\byebye{\bigskip\typeset
\sterne=1\ifx\speciali\undefined\else
\bigskip Special caracters created by the author
\loop\smallskip\noindent special character No\number\sterne:
\csname special\romannumeral\sterne\endcsname
\advance\sterne by 1\global\sterne=\sterne
\ifnum\sterne<11\repeat\fi
\vfill\supereject\end}
\def\typeset{\centerline{\petit This article was processed by the author
using the \TeX\ Macropackage from Springer-Verlag.}}
 
{\it Invited talk at Journ\'ees Relativistes 96, Ascona, Switzerland}

{\it To appear in Helvetica Physica Acta}

\title{Probing Density Fluctuations \tbreak at Low and High Redshift}
\authors{Ofer Lahav}
\address{Institute of Astronomy, Madingley Road, Cambridge CB3 0HA, UK}
\abstract{ 
We discuss cosmological inference from galaxy surveys at low and high
redshifts.  Studies of optical and IRAS redshift surveys with median
redshift ${\bar z} \sim 0.02$ yield measurements of the density
parameter $\Omega$ and the power-spectrum of density fluctuations, but
both cannot easily be related to the properties of the underlying mass
distribution due to the uncertainty in the way galaxies are biased
relative to the mass distribution.  Moreover, currently little is
known about fluctuations on scales intermediate between local galaxy
surveys ($\sim 100 h^{-1} $ Mpc) and the scales probed by COBE ($\sim
1000 h^{-1} $ Mpc).  We focus here on several issues, as examples
of clustering on different scales: the extent of the
Supergalactic Plane, an optimal reconstruction method of the density
and velocity fields, the effect of biasing on determination of
$\Omega$ from redshift distortion, the future big surveys SDSS and 2dF
(with median redshift ${\bar z} \sim 0.1$) and radio sources and the
X-Ray Background as useful probes of the density fluctuations at higher
redshift (${\bar z} \sim 1$).
}

\section{1 Introduction}

It is believed by most cosmologists that 
on the very large scales the universe obeys
the equations of General Relativity for an isotropic and homogeneous system, 
and that the Friedmann-Robertson-Walker  (FRW)  metric is valid.
However, on scales much smaller than the horizon the distribution 
of luminous matter is clumpy.
Galaxy surveys in the last decade have provided a major tool for
cosmographical and cosmological studies.  In particular, surveys such
as CfA, SSRS, IRAS, APM and Las Campanas yielded useful information on
local structure and on the density parameter  $\Omega$ from
redshift distortion and from comparison with the peculiar velocity
field.  Together with measurements of the Cosmic Microwave Background
(CMB) radiation and gravitational lensing the redshift surveys provide
major probes of the world geometry and the dark matter.

In spite of the rapid progress
there are two gaps in our current understanding of 
the density fluctuations as a function of scale:
(i) It is still unclear how to relate the distributions of light 
    and mass, in particular how to match 
    the clustering of galaxies with the CMB fluctuations,
(ii) Currently little is known about fluctuations 
on  intermediate scales 
between these of local galaxy surveys ($\sim 100 h^{-1} $ Mpc)
and the scales probed by COBE ($\sim 1000 h^{-1} $ Mpc). 

A major unresolved issue is the value of the density parameter
$\Omega$.  Putting together different cosmological observations, the
derived values seem to be inconsistent with each other.  
Taking into account moderate
biasing, the redshift and peculiar velocity data on large scales yield
$\Omega \approx 0.3 -1.5$, with a trend towards the popular 
value $\sim 1$
(e.g. Dekel 1994; Strauss \& Willick 1995 for summary of results).
On the other hand, the high fraction of baryons in clusters, combined
with the baryon density form Big Big Nucleosynthesis suggests $\Omega
\approx 0.2$ (White et al. 1993).  Moreover, an $\Omega=1$ universe is
also in conflict with a high value of the
Hubble constant ($H_0 \approx 80$ km/sec/Mpc),
as in this model the universe turns out to be younger
than globular clusters.  A   way out of these problems 
was suggested by 
adding a positive cosmological constant, such that $\Omega + \lambda
=1$, to satisfy inflation.  But two recent observations argue against
$\lambda > 0$ : the observed frequency of lensed quasars is too small,
yielding an upper limit $\lambda <0.65$ 
(e.g. Kochanek 1996 and earlier references
therein), and the
magnitude-redshift relation for Supernovae type Ia 
(Perlmutter et al. 1996)  is
consistent with deceleration parameter $q_0 = \Omega/2 - \lambda > 0
$.  So it seems that at present we are far away from knowing the world
geometry and the matter content of the universe.  The next decade will
see several CMB experiments (COBRAS/SAMBA, MAP, VSA) which promise to
determine (in a model-dependent way) the cosmological parameters to
within  a few percent.

We shall focus here on several issues related to clustering and
cosmological parameters: the extent of the Supergalactic Plane, 
an optimal  reconstruction method of the density and velocity
fields,  the
effect of biasing on determination of $\Omega$ from redshift
distortion, 
the future big surveys (SDSS and 2dF) and radio sources and
the X-Ray Background as probes of the density fluctuations at high
redshift.

\section{2  Cosmological Inference from  Redshift Surveys}
        
Redshift surveys have been used in two ways: (i) To
study the local cosmography (e.g. clusters, superclusters and voids)
and (ii) To infer statistically cosmological parameters such as the
density parameter $\Omega$ and the power-spectrum of density
fluctuations $P(k)$ (i.e. the square of the Fourier components).  These
issues were reviewed in detail by Dekel (1994) and Strauss \& Willick
(1995).  Here we shall illustrate these topics
with some specific  examples: the reconstruction of the
Supergalactic Plane (i.e. cosmography) and of the determination of
$\Omega$ (i.e. statistics), with special emphasis on the issue of
biasing.


\subsection{2.1 The Supergalactic Plane Revisited}

The so-called Supergalactic Plane (SGP) was recognized by de
Vaucouleurs (1956) using the Shapley-Ames catalogue, following an
earlier analysis of radial velocities of nearby galaxies which
suggested a differential rotation of the `metagalaxy' by Vera Rubin.
This remarkable feature in the distribution of nebulae was in fact
already noticed by William Herschel more than 200 years ago.
Tully (1986) claimed the flattened distribution of
clusters extends across a diameter of $\sim 0.1 c$ with axial ratios
of 4:2:1.  Shaver \& Pierre (1989) found that radio galaxies are more
strongly concentrated to the SGP than are optical galaxies, and that
the SGP as represented by radio galaxies extends out to redshift $z
\sim 0.02$.  
Traditionally the Virgo cluster was regarded as the centre of the
Supergalaxy, and this was termed the `Local Supercluster' .  But
recent maps of the local universe indicate that much larger clusters,
such as the Great Attractor and Perseus-Pisces on opposite sides of
the Local Group are also major components of this planar structure.
Supergalactic coordinates are commonly used in extragalactic studies,
but the degree of linearity, extent and direction of the SGP have
been little quantified in recent years.  Moreover, it is important to
compare the extent of the SGP with other filamentary structures seen
in redshift surveys and in $N$-body simulations. 
Top-down structure formation scenarios
(e.g. Hot Dark Matter) in particular predict the formation of
Zeldovich pancakes, although these are also seen in hierarchical
(bottom-up) scenarios (e.g. Cold Dark Matter).

The Optical Redshift Survey (ORS, Santiago et al. 1995) and 
the IRAS 1.2Jy survey (Fisher et al. 1995a)
have been used recently to revisit the SGP feature (Lahav et al., 
in preparation).
To estimate objectively the extent of the SGP, let us consider 
a slab embedded in a uniform sphere of radius $R$ and then construct 
the `moment of inertia tensor') for the {\it
fluctuation} in the density field:
$$
{\tilde C_{ij} } =   
{1 \over V} \sum_{gal} w_{gal}
(x_i - {\bar x_i} ) (x_j - {\bar x_j} ) -
\delta_{ij}^K {n_{bg} \over {\langle n \rangle} } { R^2 \over 5 } \;,
\eqno (1)
$$ 
where $V$ is the volume of the sphere, and $w_{gal}$ is the weight
per galaxy, which corrects for the radial and angular selection
functions of the catalogues.  $x_i, x_j$ ($i,j=1,2,3$) are Cartesian
components of ${\bf x}$, $n_{bg}$ is the {\it background} density in
the absence of the slab, and $\langle n \rangle $ 
the mean density ({\it including}
the slab).  The last term is due to a uniform distribution with
density $n_{bg}$.  
By diagonalising the moment of inertia it is found
that the angle 
between the normal to the
standard SGP and the normal to our objectively identified plane
is $\theta_z \approx 30^o$ out to $R=8000$ km/sec.
The
probability of these normals to be within an angle $\theta_z$ by
chance is $P(< \theta_z) = 1-\cos(\theta_z)$, i.e. $\sim 13 \%$.
Unlike the eigen-vectors, the eigen-values of eq. (1), which
correspond to the square of the 'ellipsoid' axes, depend on the
uncertain background density $n_{bg}$.  Preliminary analysis shows
that the SGX and SGY dimensions  are indeed much larger than SGZ, and they
extend out radius of at least 6000 km/sec.  It is a challenge for
any cosmological model to reproduce such a large pancake feature.

It is worth noting that one source of uncertainty in quantifying the
connectivity of the SGP is that disk of the Milky Way obscures about
20\% of the optical extragalactic sky, this is the so-called ``Zone of
Avoidance'' (ZOA).  The ZOA is nearly perpendicular to the SGP.
Galaxies behind the ZOA are difficult to detect due to  
extinction by dust and gas at optical wavelengths, and
confusion with Galactic stars.  
But they can be detected in
emission at 21cm by neutral atomic hydrogen (HI).  Recent discoveries
of galaxies hidden behind the ZOA include the Sagittarius dwarf (Ibata,
Gilmore and Irwin 1994 ), Dwingeloo1 (Kraan-Korteweg et al. 1994) and the cluster A3627
(Kraan-Korteweg et al. 1996), which is possibly at the centre of the
Great Attractor region.

\subsection{2.2 Wiener Reconstruction of IRAS }

Apart from the ZOA, two other major problems affect most analyses  of
 redshift surveys: shot noise and redshift distortion.  One approach
 to deal with both is Wiener filtering (Fisher et al 1995b).  
 Let us expand the density field $\rho(\bf r) = [1+
 \delta({\bf r})] {\bar \rho} $ in spherical harmonics and radial
 Bessel functions:
$$
\rho({\bf r} ) =
\sum_l \; \sum_{m} \; \sum_{n} C_{ln}\;
 \rho_{lmn} \;\; j_l(k_{n} r)   \;\; Y_{lm}({\bf \hat r}) \;, 
\eqno (2)
$$ 
where the $k_n$'s are chosen e.g. to satisfy the boundary condition
that the logarithmic derivative of the potential is continuous at
$r=R_{max}$.  An estimator of the coefficients from the redshift data
is
$$ 
 {\rho_{lmn}}^S = \sum_{gal} { 1 \over \phi(s)} \;
 j_l(k_{n} s)  \; Y_{lm}^*( {{\bf \hat r}}) \;,
\eqno (3)
$$ 
where $\phi(s)$ is the radial selection function.
Fisher et al. (1995b) showed that the real-space coefficients 
of the fluctuations can 
be reconstructed by 
$$ 
\delta_{lmn}^R  =  \sum_{n' n''}
\left( \bf S_l \left[ \bf S_l + \bf N_l \right]^{-1}\right)_{n n'}
\left( \bf Z_l^{-1}\right)_{ n' n''} \delta_{lmn''}^S \;, 
\eqno (4)
$$ 
where the matrix ${\bf Z_l}$ (which depends on the assumed
combination of density and biasing parameters,
$\beta = \Omega^{0.6}/b$) converts the redshift space 
coefficients to real space coefficients.  $\bf S_l
\left[ \bf S_l + \bf N_l \right]^{-1}$  is the Wiener matrix, 
roughly representing signal/(signal+noise),
which
filters the data where they are noisy. The signal matrix ${\bf S_l}$
depends on the assumed prior power-spectrum of fluctuations.  It can
be shown that this gives the optimal reconstruction in the minimum
variance sense, and it can also be derived from Bayesian arguments and
Gaussian probability distribution functions.  In this approach the
density field goes to the mean density at large distances. This does
not mean necessarily that the density field itself disappears at large
distances, it only reflects our ignorance on what exists far away,
where the data are very poor.

Webster, Lahav \& Fisher (1996) have utilised this method to recover
the density and velocity fields from the IRAS 1.2Jy redshift
survey. 
Many known structures are seen in
the reconstructed maps, including clear confirmations of the clusters
N1600 and A3627. The Perseus-Pisces supercluster appears to extend out
to roughly 9000 km/sec, and the reconstruction shows `backside infall' to
the Centaurus/Great Attractor region. 
The
Wiener reconstruction of the density field is also the optimal
reconstruction (in the minimum variance sense) of any quantity which
is linear in the density contrast such as the dipole, bulk flows and
the SGP.  The misalignment angle between the IRAS and CMB Local Group
dipoles is only 13 degrees out to 5000 km/sec, but increases to 25
degrees out to 20,000 km/sec (see Figure 1). 
It still has to be understood if this misalignment 
is due to non-linear effects or missing gravity due to the ZOA, 
power on large scales and/or biasing.
The reconstructed IRAS bulk flow out to
5000 km/sec is roughly 300 km/sec (for $\beta=0.7$),  
similar in amplitude to  that
derived from the Mark III peculiar velocities (370 km/sec). However, the
two bulk flow vectors deviate by 70 degrees. Finally,  a moment of
inertia analysis shows that the Wiener reconstructed SGP
is aligned within 30 degrees of that defined by de Vaucouleurs,
in agreement with the analysis discussed in the previous section.

\begfig 3 cm
\vskip -3 cm 
\figure{1}{The direction of  the Wiener reconstructed 
IRAS dipole compared with the COBE dipole (in the Local Group 
frame). The reconstruction assumes 
the observed power-spectrum of IRAS galaxies 
(characterized by $\Gamma$ and $\sigma_8$)
and $\beta=\Omega^{0.6}/b=0.7$.
The crosses show the convergence of the direction of the 
reconstructed IRAS dipole. Starting at the top of the plot, the crosses
give the direction (in Galactic coordinates) of the dipole within 
radius $R$ (in steps of 1 $h^{-1}$ Mpc).
The circular curves denote separations from the COBE result in $10^o$ 
intervals. From Webster, Lahav \& Fisher (1996)}
\endfig


\section{3 $\Omega$ and Bias Parameter from Redshift Distortion }

In the previous section we used IRAS galaxies to represent the sources
of gravity in the local universe. 
However, it is most likely that luminous galaxies do not trace perfectly 
the mass distribution. This effect is commonly phrased as 'biasing',
although there is some confusion (or over-simplicity) in modelling it.
Kaiser (1984) formulated biasing 
in terms of statistics of peaks. He showed that in the linear approximation 
the correlation function of galaxies is related to 
the mass correlation function by 
$$
\xi_{gg} = b^2 \xi_{mm}
\eqno (5)
$$
where $b$ is the 'bias parameter'. It has become a common practice
to assume that the galaxy  and mass 
density fluctuations at any point are related by
$$
\delta_g = b \delta_m 
\eqno (6)
$$
Clearly , if eq. (6) is true then eq. (5) follows, but
the reverse is not true.
Usually, eq. (6) is assumed in different statistics, 
but if it does not hold (which is very likely the case), 
then one compares `apples and oranges' in the various determinations
of $\Omega$.

An illustration that biasing can be more complicated than eq. (6) 
is given when the clustering properties of different galaxy 
morphologies are compared.
For example, Hermit et al. (1996) found that the 
relative bias factor between early type galaxies
and late-types  weakly depends on scale out to 10 $h^{-1}$ Mpc.
On the other hand, on much larger scales voids seem empty for 
all known galaxy populations. It is therefore crucial to generalize eq. 
(6) for more realistic scenarios.
Many extensions for biasing are possible:
non-linear, non-local,  scale-dependent, epoch-dependent and stochastic
(Dekel \& Lahav, in preparation).
Here we give a specific example on how the determination 
of $\Omega$ from redshift distortion could  be modified.

\subsection{3.1 Biasing in Redshift Distortion  }

Studies of redshift distortion in galaxy redshift surveys 
aim at deducing $\beta = \Omega^{0.6}/b$.
Derived values by this method cover the range $0.45 < \beta < 1.10$
(Strauss \& Willick 1995).
However, the galaxies play two roles 
in such analysis: they are both luminous tracers of the mass distribution 
as well as test particles of the velocity field,
and hence the form of biasing is more complicated.

Kaiser (1987) showed that in linear theory and in the far field
approximation  the density fluctuation in galaxies in redshift space 
$\delta_{g,S}$ 
is related to the one in real space 
$\delta_{g,R}$ 
by 
$$
\delta_{g,S} = \delta_{g,R} - { dU \over dr}
\eqno (7)
$$
where ${ dU \over dr} = -\mu^2 \Omega^{0.6} \delta_m$ 
is the gradient of line of sight velocity in linear theory 
and $\mu$  is the cosine of the angle between the line of sight and the 
${\bf k}$ vector.
We then find  that the power spectrum in redshift space
is:
$$
P_{gg}^S(k) = P_{gg}^R(k) + 2 P_{mg}^R(k) \Omega^{0.6} \mu^2 +   
\Omega^{1.2} \mu^4 P_{mm}^R (k) 
\eqno (8)
$$
where $P_{gg}^R(k), P_{mg}^R(k) $ and $P_{mm}^R(k)$ 
are the galaxy-galaxy, mass-galaxy and mass-mass power spectra
in real space.
This generalizes eq. (3.5) of Kaiser (1987).
Only if $P_{gg}^R(k) = b^2 P_{mm}^R(k) $ and 
$P_{gg}^R(k) = b P_{mg}^R$ the relation goes back to the 
simple and much-used relation: 
$$
P_{gg}^S(k) = P_{gg}^R(k) ( 1 + \beta \mu^2 )^2. 
\eqno (9)
$$

Similarly, the spherical harmonic analysis 
for redshift distortion in linear theory 
for a flux-limited survey 
(Fisher, Scharf \& Lahav 1994; eq. 11)
can be extended to give for the 
the mean-square predicted harmonics: 
$$
\langle | a_{lm}^S |^2 \rangle =
 \, {{2}\over{\pi}}  \;\int
dk\,k^2 \{ P_{gg}^R(k) 
\left| \Psi_l^R (k)\right|^2  
+ 2 \Omega^{0.6} P_{mg}^R (k) \left| \Psi_l^R (k) \Psi_l^C(k)\right| \; 
+ \Omega^{1.2} P_{mm}^R(k) \left| \Psi_l^C (k)\right|^2 \}  
\eqno (10)
$$
where $\Psi_l^R(k)$ 
and $\Psi_l^C(k)$ are the real space and redshift correction 
window functions which depend on the selection and weighting functions.
We see that if the density fields of mass and  light do not obey 
linear biasing then  
direct comparison with $\beta$ derived by other methods 
is inconsistent. Better modelling of biasing may help resolving the 
discrepancies between the different values obtained for $\Omega$.


\section{4 Future  Redshift Surveys: SDSS  and 2dF }

Existing optical and IRAS redshift surveys contain 10,000-20,000
galaxies.  A major step forward using multifibre technology will allow
in the near future to produce redshift surveys of millions of
galaxies.  In particular, there are two major surveys on the horizon.
The American-Japanese 
 Sloan Digital Sky Survey (SDSS) will yield images in 5
colours for 50 million galaxies, 
and redshifts for about 1 million galaxies over a quarter of the
sky (Gunn and Weinberg 1995). It will
be carried out using a dedicated 2.5m telescope in New Mexico.  The
median redshift of the survey is $z \sim 0.1$.
A complementary Anglo-Australian survey, the 2 degree Field (2dF), will
produce redshifts for 250,000 galaxies brighter than $b_J =19.5^m$
(with median redshift of $z \sim 0.1$), selected from the APM catalogue.  The
survey will utilize a new 400-fibre system on the 4m AAT, covering
$\sim 1,700$ sq deg of the sky.  
About 250,000 spectra will be measured over $\sim
100$ nights.
A deeper
extension down to $R=21$ for 10,000 galaxies is also planned for the 2dF 
survey.
These  surveys will probe
scales larger than $\sim 30 h ^{-1}$ Mpc.
It will also allow accurate determination
of $\Omega$ and bias parameter from redshift distortion.  
Surveys like 2dF and SDSS will produce  unusually large numbers of galaxy
spectra, providing an important probe of the intrinsic
galaxy properties,  for studying e.g. the density-morphology relation. 
Several groups 
(e.g. Connolly et al. 1995 ;  Sodr\'e
\& Cuevas 1996; Folkes, Lahav \& Maddox 1996) 
recently devised techniques for automated spectral classification
of galaxies.
These techniques include e.g. Principal Component Analysis 
and Artificial Neural Networks.

\section{5 Probes of density fluctuations at high redshift }

The big new surveys 
(SDSS, 2dF) 
will only probe a median redshift ${\bar z} \sim 0.1$.
It is still crucial to probe the density fluctuations at higher $z$,
and  to fill in the gap between
scales probed by previous local galaxy surveys and the scales 
probed by COBE and other CMB experiments.
Here we discuss the X-ray Background (XRB) and radio sources 
as probes of the density fluctuations at median redshift $ z \sim 1$.
Other possible high-redshift traces are quasars and clusters of galaxies.
For review on the evolution of galaxies with redshift see e.g. 
Fukugita, Hogan \& Peebles (1996).  

\subsection{5.1 The X-ray Background }

Although discovered in 1962, the origin of
the X-ray Background (XRB) is still unknown,  
but is likely
to be due to sources at high redshift 
(for review see Boldt 1987; Fabian \& Barcons 1992).
Here we shall not attempt to speculate on the nature of the XRB sources.
Instead, we {\it utilise} the XRB as a probe of the density fluctuations at
high redshift.  The XRB sources are probably
located at redshift $z < 5$, making them convenient tracers of the mass
distribution on scales intermediate between those in the CMB as probed
by COBE, and those probed by optical and IRAS redshift
surveys (see Figure 2).
  In terms of the level of anisotropy, the XRB is
also intermediate between the tiny CMB fluctuations ($\sim 10^{-5}$ on angular
scales of degrees) and galaxy density fluctuations (of the order of unity on
scale of 8 $h^{-1}$ Mpc).

The preliminary measurements of the dipole anisotropy in the XRB
(Shafer 1983) were discussed qualitatively (e.g. Rees 1979) 
by associating it with
local clusters such as Virgo and the Great Attractor and by other
cosmographical arguments. 
Lahav, Piran \& Treyer (1996) recently treated the problem in a
statistical rather than cosmographical way.  They predicted rms
spherical harmonics in the framework of growth of structure by
gravitational instability from density fluctuations drawn from a
Gaussian random field.  The XRB harmonics are expressed in terms of
the power-spectrum of density fluctuations and for evolution scenarios
which are consistent with recent measurements of galaxy clustering and
the CMB.  The dipole is due to large scale structure as well as to the
observer's motion (the Compton-Getting effect).  For a typical
observer the two effects turn out to be comparable in amplitude.  The
coupling of the two effects makes it difficult to use the XRB for
independent confirmation of the CMB dipole being due to the observer's
motion.  The large scale structure dipole (rms per component) relative
to the monopole is in the range $a_{1m}/a_{00} \sim (0.5-9.0) \times
10^{-3} $.  The spread is mainly due to the assumed redshift evolution
scenarios of the X-ray volume emissivity $\rho_x(z)$.  The dipole's
prediction is consistent with a measured dipole in the HEAO1 XRB map.
Typically , the harmonic spectrum drops with $l$ like $a_{lm} \sim
l^{-0.4}$.  This behaviour allows us to discriminate a true clustering
signal against the flux shot noise, which is constant with $l$, and
may dominate the signal unless bright resolved sources are removed
from the XRB map.  The Sachs-Wolfe and Doppler (due to the motion of
the sources) effects in the XRB are negligible.
Measurements of the spherical harmonic spectrum in maps such as HEAO1 and 
ROSAT could provide important constrains on amount of clustering
at high redshift.

\begfig 3 cm
\vskip -3 cm 
\figure{2}{The quadrupole ($l=2$) `window functions' in Fourier space
for the CMB(Sachs-Wolfe effect), the X-ray Background (for specific model parameters), 
and IRAS galaxies.  
The hight  of the window functions is arbitrary. 
The solid and dashed lines represent 
$k^3 P(k) \sim (\delta \rho /\rho)^2 $
for standard Cold Dark Matter model and the observed galaxy power spectrum 
(fitted by low density CDM model), respectively.
 From Lahav, Piran \& Treyer (1996).  
} 
\endfig

\subsection{5.2 Radio Sources }

Surveys of radio sources have typically median redshift ${\bar z} \sim
1$, and hence are useful probes of clustering at high redshift.
Unfortunately, it is difficult to obtain distance information from
these surveys: the radio luminosity function is very broad, and it is
difficult to obtain optical redshifts of the radio sources.

Several groups (Kooiman et al 1995, Sicotte 1995, Loan, Wall and
Lahav 1996, Cress et al. 1996) have recently studied clustering of
radio sources.  For example, Loan et al. (1996) measured the angular
two-point correlation function $w(\theta)$ in the Green Bank and
Parkes-MIT-NRAO 4.85 GHz surveys.  The signal is noisy, but with an
assumed redshift distribution indicates strong clustering in 3
dimensions.  It is convenient to parameterize the evolution of the
spatial correlation function in comoving coordinates as:
$$
\xi(r_c,z) =
( r_c / r_0 )^{-\gamma} \; (1+z)^{\gamma-(3+\epsilon)}.
\eqno (11)
$$
For
$\gamma=1.8$,
and `stable clustering' ($\epsilon=0$)
the derived correlation length is $r_0 \approx 18 h^{-1}$ Mpc, larger than the
value for nearby normal galaxies and comparable to the cluster-cluster
correlation length. This may suggest that radio sources are associated with 
high-density regions. This is in accord with earlier studies 
(Bahcall \& Chokshi 1992, Peacock \& Nicholson  1991) 
and the new studies mentioned above.
It is of interest to detect the dipole and higher harmonics in these
radio surveys. As for the XRB, it is found that the motion 
(Ellis \& Baldwin 1984)
and large scale structure dipole effects are comparable
(Baleisis 1996), but both are smaller than the shot noise. 
New surveys, such as FIRST and NVSS will  constrain much better 
the clustering properties at high redshift.

\section{6 Discussion }

We have shown some recent studies 
of  galaxy  surveys, and their cosmological implications. 
Local IRAS and optical surveys have been used to describe
the local cosmography (e.g. the Supergalactic Plane)
and to constrain $\Omega$ (e.g. from redshift distortion and dipoles).
However, the issue of biasing is still conceptually underdeveloped, 
and is crucial for analysing the new big surveys (2dF, SDSS).
To study galaxy evolution and the validity 
of the FRW metric on large scales 
it is important to explore density fluctuations 
at higher redshift. The examples of the X-ray Background and radio sources
are encouraging, but redshift information is required to constrain 
the growth of cosmic structure with time.
With the dramatic increase of data, we should soon be able to map
the fluctuations with scale and epoch.

{ \bf Acknowledgments} 
I thank my collaborators for their contribution to the work
presented here, and for stimulating  discussions,
and the organizers of JR96  for an enjoyable meeting. 
\section{References}
                                
\countdef\refno=30
\refno=0
\def\ref{\advance \refno by 1 \ifnum\refno<10 \item{ [\the\refno ]}
\else \item{[\the\refno ]} \fi}
\outer\def\section#1\par
           {\advance \sectno by 1
           \vskip0pt plus .3\vsize\penalty-250\vskip 0pt plus-.3
           \vsize \bigskip\vskip\parskip
           \noindent\leftline{\rlap{\bf \the\sectno .}
           \hskip 17pt
            \underbar{\bf #1\kern-4pt}}\nobreak\smallskip}
\ref       Bahcall N.A. \&  Chokshi, A., 1992, ApJ,  382, L33
\ref       Baleisis A., 1996, M.Phil. thesis, Cambridge University
\ref       Boldt E., 1987, Phys. Reports, 146, 215
\ref       Connolly, A.J., Szalay, A.S., Bershady, M.A., 
           Kinney, A.L. \& Calzetti, D.
           1995, AJ, 110, 1071
\ref       Cress C.M., Helfand D.J., Becker R.H., Gregg M.D., White R.L.,
           1996, ApJ, in press 
\ref       Dekel, A., 1994, ARA\&A, 32, 371
\ref       de Vaucouleurs, G., 1956, Vistas in Astronmy, 2, 1584
\ref       Ellis, G.F.R. \& Baldwin, J.E. 1984, MNRAS, 206, 377
\ref       Fabian A.C. \&  Barcons X., 1992, ARA\&A, 30, 429
\ref       Fisher, K.B., Scharf, C.A. \& Lahav, O., 1994, 266, 219
\ref       Fisher, K.B., Huchra, J.P, Davis, M., Strauss, M.A.,
           Yahil, A., \& Schlegel, D. 1995a, ApJS, 100, 69
\ref        Fisher, K.B., Lahav, O., Hoffman, Y., Lynden-Bell, D., 
           \& Zaroubi, S. 1995b,  MNRAS, 272, 885 
\ref       Folkes, S., Lahav, O. \& Maddox, S.J., 1996, 
           MNRAS, in press 
\ref       Fukugita, M., Hogan, C.J. \& Peebles, P.J.E., 1996, Nature, 381, 489
\ref       Gunn, J.E. \& Weinberg, D.H., 1995, 
           in {\it Wide-Field Spectroscopy and the Distant Universe}, 
           eds. S.J. Maddox \& A. Aragon-Salamanca, World Scientific, 
           Singapore 
\ref       Hermit, S. Santiago, B.X., Lahav, O., Strauss, M.A.,
           Davis, M., Dressler, A., \& Huchra, J.P., 1996, MNRAS, in press 
\ref       Ibata, R.A., Gilmore, G. \& Irwin, M.J., 1994, Nature, 370, 1941
\ref       Kaiser, N., 1984, ApJ,  284, L9
\ref       Kaiser, N., 1987, MNRAS, 227, 1 
\ref       Kochanek, C.S., 1996, ApJ, 466, 638
\ref       Kooiman B.L., Burns J.O., Klypin A.A., 1995, ApJ, 448, 500
\ref       Kraan-Korteweg, R.C.,  Loan, A.J.,  Burton, W.B.,  Lahav, O., 
           Ferguson, H.C.,  Henning, P.A., \&   Lynden-Bell, D., 1994,
           Nature, 372, 77
\ref       Kraan-Korteweg, R.C.,  Woudt, P.A., Cayatte, V., Fairall, A.P., 
           Balkowski, C. \& Henning, P.A., 1996, Nature, 379, 519
\ref       Lahav, O., Piran, T. \& Treyer M.A.,  1996, MNRAS, in press 
\ref       Loan, A.J., Wall, J.V. \& Lahav, O., 1996, MNRAS, submitted
\ref       Peacock J.A \&  Nicholson D., 1991, MNRAS, 253, 307
\ref       Perlmutter, S., et al., 1996, preprint
\ref       Rees, M.J., 1979, {\it Objects of High Redshifts}, IAU Symp.,
           G. Abell \& J. Peebles Eds.
\ref       Santiago, B.X., Strauss, M.A., 
           Lahav, O., Davis, M., Dressler, A., \& Huchra, J.P. 1995, 
           ApJ, 446, 457
\ref       Shafer, R.A., 1983, PhD Thesis,  
           University of Maryland, NASA TM-85029 
\ref       Shaver, P.A., \& Pierre, M., 1989, A\&A, 220, 35
\ref       Sicotte H., 1995, Ph.D. thesis, Princeton University
\ref       Sodr\'e L. Jr. \& Cuevas, H., 
           1996, MNRAS, submitted 
\ref       Strauss, M.A. \& Willick, J.A. 1995, Phys. Reports, 261, 271
\ref       Tully, B.R., 1986, ApJ, 303, 25
\ref       Webster, A.M., Lahav, O. \& Fisher, K.B., 1996, MNRAS, submitted
\ref       White, S.D.M., Navarro, J.F., Evrard, A.E. \& Frenk, C.S.,
           1993, Nature, 366, 429
\end